\documentstyle[preprint,revtex]{aps}
\input{epsf}
\begin{document}
\draft
\preprint{SLAC-PUB-5968}
\preprint{November 1992}
\preprint{T}
\begin{title}
Nucleon Form Factors in a Relativistic Quark Model \footnote{Work supported
in part by the Department of Energy, contract DE-AC03-76SF00515.}
\end{title}
\author{Felix Schlumpf}
\begin{instit}
Stanford Linear Accelerator Center\\
Stanford University, Stanford, California 94309
\end{instit}
%\receipt{}
\begin{abstract}
We demonstrate that a relativistic constituent-quark
model can give nucleon form factors in agreement with data for
both low and high momentum transfer. The relativistic features
of the model and the specific form of the wave function are essential
for the result.
\end{abstract}
\narrowtext

The purpose of this report is to show that a simple constituent-quark model
can yield the elastic nucleon form factors in agreement with all available
data up to more than 30 GeV$^2$. Very high momentum transfer ($Q^2$)
behavior of elastic form factors can be obtained from
perturbative QCD \cite{farrar79}, while
low energy quantities are calculated within various models. But
there is still an open question of the energy scale at which the
perturbative contributions are important. A model analysis of the
pion \cite{kisslinger} concluded that the nonperturbative contributions are
much larger at 2 GeV$^2$. An analysis based on general features
of models \cite{isgur} found also for the nucleon that perturbative terms are
unimportant in the region of present experiments.
It is therefore important to
have a model that is valid for all values of $Q^2$.

The hadronic matrix element of the radiative transition of the nucleon
$N \to N' \gamma$ is represented in terms of the form factors as
\begin{eqnarray}
\lefteqn{\left< N', p' \left| J^\mu \right| N, p\right>  =}\nonumber\\
&&\bar u(p') \left[ F_1(Q^2)\gamma^\mu + {F_2(Q^2) \over
2 M_N}i\sigma^{\mu\nu}Q_\nu \right] u(p) ,
\end{eqnarray}
with momentum transfer $Q=p'-p$, nucleon mass $M_N$, and quark current
$J^\mu =\bar q \gamma^\mu q$. The Sachs form factor for the
magnetic transition is given by $G_M=F_1+F_2$. The matrix
elements can be calculated within a relativistic constituent quark model
on the light cone \cite{terentev,aznauryan,coester91}.
This approach has been extended to asymmetric
wave functions \cite{schlumpf},
which provide an excellent and consistent picture
of electroweak transitions of the baryon octet. In this report we
focus on the high energy behavior of the wave function.

Usually harmonic-oscillator-type wave functions are used
\cite{terentev,aznauryan,coester91}
\begin{equation}
\phi (M) =N \exp \left( -M^2/2\alpha^2 \right) ,
\end{equation}
with $\alpha$ being the confinement scale of the bound state and
$N$ being the normalization. The operator $M$ is the free mass
operator of the noninteracting three-body system, and it is a function
of the internal momentum variables $\vec q_i$ of the quarks
and the quark mass $m$:
\begin{equation}
M=\sum_i \sqrt{\vec q_i\,^2 + m^2}.
\end{equation}
With this special form of the wave function the form factors fall off
exponentially for high $Q^2$. This is why the form factors calculated with
Eq.~(2) are only valid up to 4--6 GeV$^2$, an energy scale well below the
perturbative region \cite{isgur}.

The orbital wave function we use is
\begin{equation}
\phi (M)=\frac{N}{(M^2+\alpha^2)^{3.5}} ,
\end{equation}
with a scale $\alpha$ and normalization factor $N$ different
from Eq.~(2). The two parameters of the model, the confinement scale
$\alpha$ and the quark mass $m$, have to be determined by comparison
with experimental data. We find a quark mass $m=263$ MeV and a scale
$\alpha=607$ MeV by fitting the magnetic moment of the proton $\mu(p)$
and the neutron $\mu(n)$.
In addition, these parameters give also excellent results for the
magnetic moments and the semileptonic decays of the baryon
octet \cite{schlumpf}.

 For reference, we calculate
the form factors with Eq.~(2)
using parameters $\alpha=560$ MeV and $m=267$ MeV as well.
The results are shown in Figs. 1--3. Note that the low energy behavior of
both wave functions is almost identical, while only the wave function in
Eq.~(4) fits the data. It is therefore significant to choose
an appropriate Ansatz for the orbital wave function.
The relativistic features of the model are also
important. In the nonrelativistic limit, $\alpha / m \to 0$, the form
factor $G_M$ for the proton is (for small Q$^2$)
\begin{equation}
\frac{G_M}{\mu(p)}=\cases{\exp\left( -Q^2/\alpha^2\right)
                            &[Eq.~(2)] ,\cr
              \left( 1+ \frac{2 Q^2}{\alpha^2+9 m^2}\right)^{-3.5}
			               &[Eq.~(4)] ,}
\end{equation}
which is too small for any reasonable value of $\alpha$ and
$m$ (compare Figs. 1 and 2).
This limit shows that the relativistic treatment of the problem increases
 the form factors significantly, even for low values of the momentum
transfer. The same effect has also been found for the pion \cite{coester88}.
While the asymptotic falloff for the wave function in Eq.~(4) is
still larger than $Q^{-4}$, it shows up only at very high
$Q^2$ of over 1000 GeV$^2$.

We conclude that quark models with reasonable parameters can give
agreement with data for all $Q^2$. The nonrelativistic limit is not
adequate, and the specific form of the wave function is essential.

\acknowledgments

I would like to thank W.~Jaus for helpful discussions.
This work was supported in part by the Schweizerischer Nationalfonds and
in part by the Department of Energy, contract DE-AC03-76SF00515.

\figure{The proton form factor $F_1(Q^2)$: continuous line, Eq.~(4);
broken line, Eq.~(2). The experimental points are taken from
Ref.~\cite{exp}.\label{autonum}}

\figure{The proton form factor $F_2(Q^2)$: continuous line, Eq.~(4);
broken line, Eq.~(2). The experimental points are taken from
Ref.~\cite{exp}.}

\figure{The proton form factor $G_M(Q^2)$: continuous line, Eq.~(4);
broken line, Eq.~(2). The experimental points are taken from
Ref.~\cite{arnold86}.}

\newpage
\centerline{\epsfysize=\hsize \epsfbox{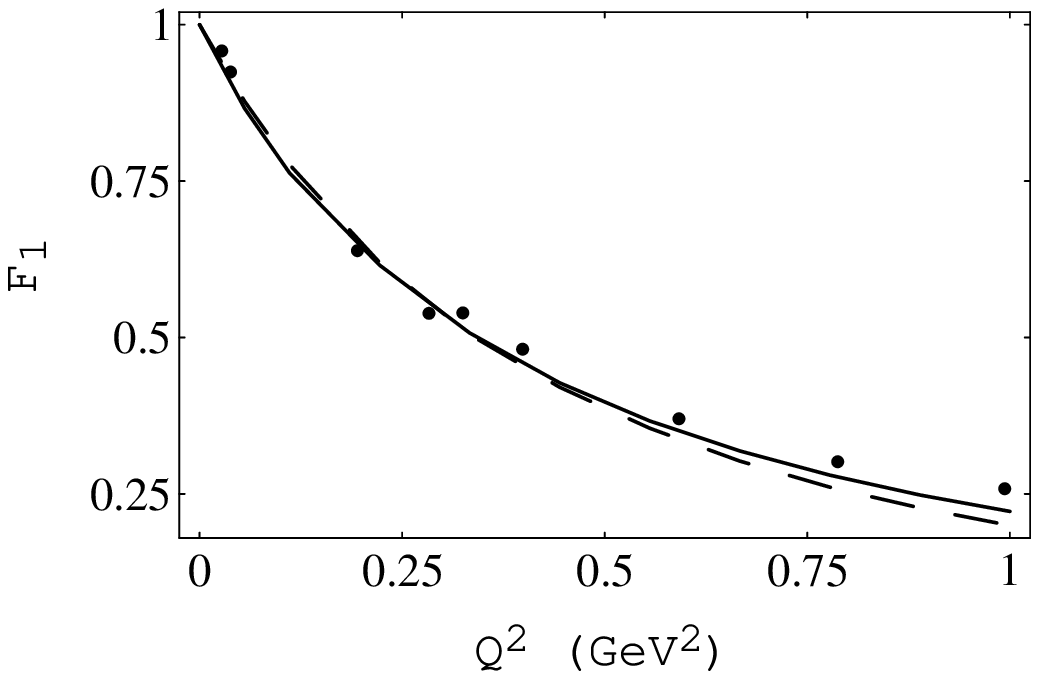}}
\vskip 3cm\centerline{Fig. 1}\newpage
\centerline{\epsfysize=\hsize \epsfbox{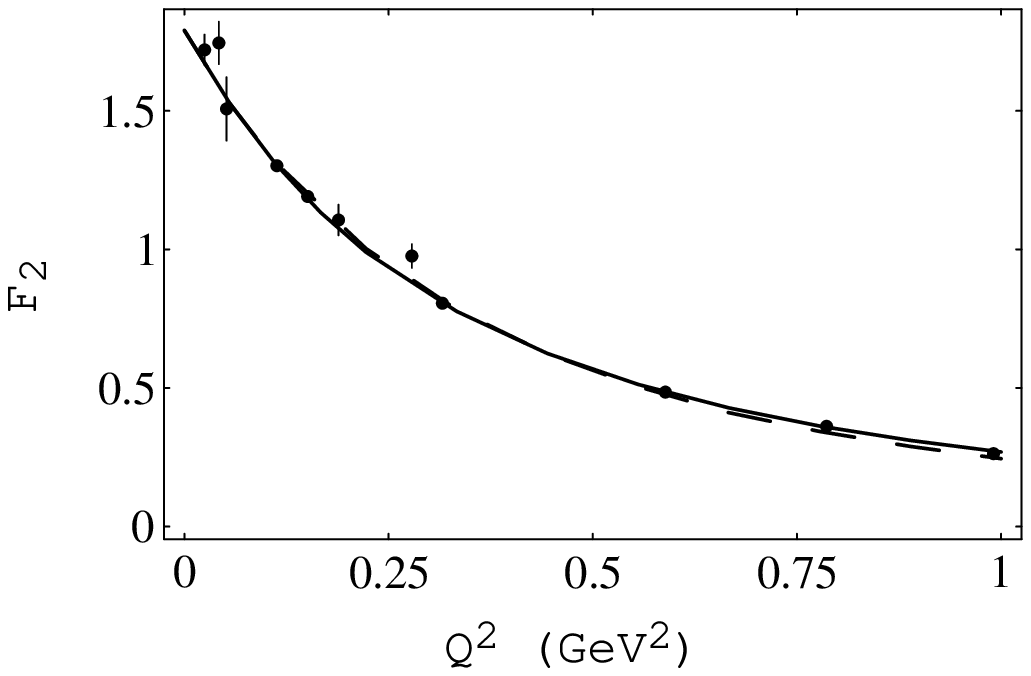}}
\vskip 3cm\centerline{Fig. 2}\newpage
\centerline{\epsfysize=\hsize \epsfbox{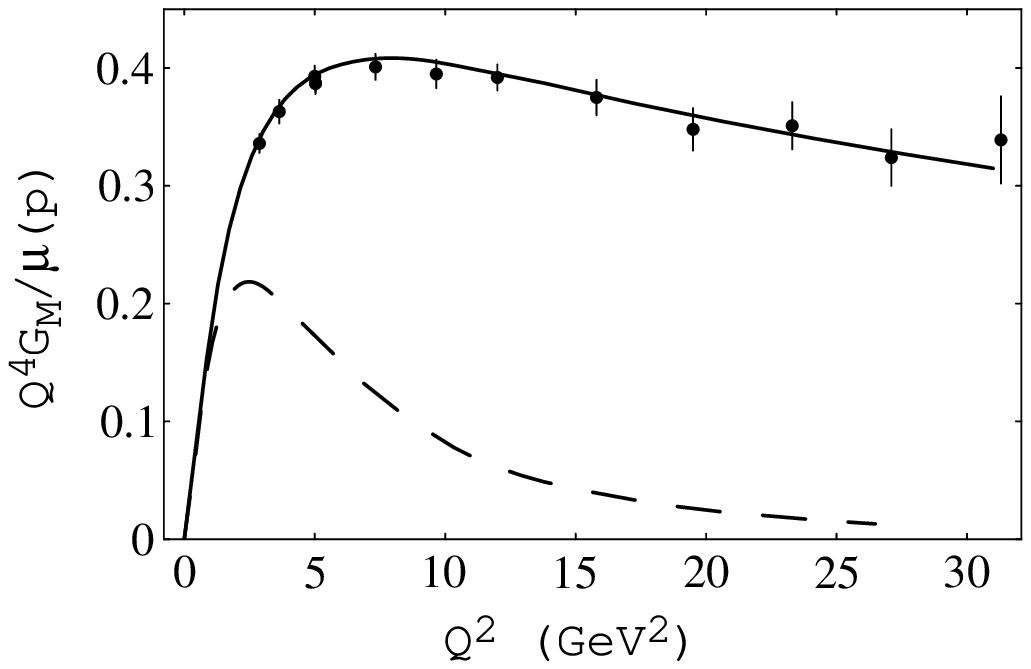}}
\vskip 3cm\centerline{Fig. 3}

\end{document}